\documentclass[reprint,
 amsmath,amssymb,
 aps,
pra,
]{revtex4-2}

\usepackage{graphicx}% Include figure files
\usepackage{dcolumn}% Align table columns on decimal point
\usepackage{bm}% bold math
\usepackage[caption=false]{subfig}
\usepackage{xcolor}
\usepackage{multirow}

\makeatletter

\def\@seccntformat#1{\csname the#1\endcsname.\quad}
\makeatother

\begin{document}

%\preprint{APS/123-QED}

\title{Insights into the adhesion and delamination strength of carbon films on metals by high-throughput \textit{ab initio} calculations}

\author{Elisa Damiani}%
\affiliation{ Department of Physics and Astronomy, University of Bologna, Viale Carlo Berti Pichat 6/2, Bologna, 40127, Italy}
\author{Margherita Marsili}%
\affiliation{ Department of Physics and Astronomy, University of Bologna, Viale Carlo Berti Pichat 6/2, Bologna, 40127, Italy}
%\email{margherita.marsili@unibo.it}

\author{M. Clelia Righi}%
\email{clelia.righi@unibo.it}
%\homepage{https://tribchem.it/}
\affiliation{ Department of Physics and Astronomy, University of Bologna, Viale Carlo Berti Pichat 6/2, Bologna, 40127, Italy}
%\email{clelia.righi@unibo.it}

\date{\today}% It is always \today, today,

\begin{abstract}
Diamond and diamond-like carbon (DLC) coatings are widely employed for their exceptional mechanical, thermal and chemical properties, but their industrial application is often limited by weak adhesion to metallic substrates. In this work, we employ a high-throughput \textit{ab initio} approach to systematically investigate the adhesion of diamond/metal interfaces, combining a set of technologically relevant metals (Al, Ag, Au, Cr, Cu, Fe, Ir, Mg, Mo, Pt, Rh, Ti, V, W, Zn) with the C(111), C(111)-2$\times$1 (Pandey reconstructed), C(110), C(100) surfaces, that are most common in diamond and are representative of different types of bonds present in DLC. Thanks to our automated and accurate computational protocol for interface construction and characterization, databases are populated and relevant trends are identified on the effect of surface graphitization, ability to form carbides and metal reactivity on carbon film adhesion and delamination strength. Beyond capturing trends, our workflow yields predictive insights. Indeed, we found that adhesion energy scales with the geometric mean of the constituent surface energies, providing a simple descriptor for rapid screening; while comparing the work of separation with the metal’s cohesive energy anticipates the fracture location under tensile loading. A novel method based on the radial distribution function $g(r)$ analysis is introduced to identify when contact with a metal drives rehybridization of surface carbon from sp$^2$ to sp$^3$, the structural signature of improved resistance to delamination. These structural changes are mirrored by an electronic rearrangement at the interface, quantified by a charge-redistribution descriptor that strongly correlates with adhesion.
%Adhesion energies of 60 interfaces are calculated and analyzed in relation to intrinsic surface properties of the constituents, interface fracture behaviour and electronic charge accumulation. Moreover, the structural and electronic rearrangement of diamond reconstructions upon interface formation has been discussed. Our results confirm that carbide-forming metals such as Cr, Ti, W and V adhere more strongly to diamond surfaces, while noble or less reactive metals with filled outer shells (Ag, Au, Zn, Mg) show weak bonding. At the same time, the C(111) slab, exhibiting surface sp$^3$ bonds, systematically presents higher adhesion with metal rather than C(111)-2$\times$1, where sp$^2$ carbon atoms are present at the surface. Our analysis suggests that the preservation of surface sp$^2$ bonds in C(111)-2$\times$1 strongly influences adhesion, with more reactive metals inducing partial rehybridization of the surface chains thus enhancing adhesion. 
\end{abstract}

\maketitle

\section{INTRODUCTION} \label{sec:introduction}
Diamond displays unmatchable properties: extreme hardness, chemical inertness and biocompatibility, exceptionally high thermal conductivity, low dielectric constant, high thermal stability and excellent resistance to radiation. Due to these qualities carbon-based films, made of amorphous diamond like carbon (DLC) and poly- or nano-crystalline diamond, have found application within the development of sensors, electrodes, high power and high temperature devices and substrates for biological activity able to comply also with harsh environments \cite{diamond_properties, biocompatibility_2, biocompatibility, biocompatibility_3}. Besides these qualities, diamond and DLC films coatings provide some of the lowest known wear and friction coefficients \cite{ROBERTSON2002129, LEPICKA201762} and are therefore applied in industry to minimize friction and wear in engine parts, prevent stiction in micro- and nano-electromechanical systems (MEMS and NEMS) and enhance durability and performance of industrial cutting tools and biological implants \cite{Solid_lubr, MEMS, bio_implants, DLC_wear}.\par 
Being able to control adhesion of a DLC film by tuning the parameters that govern it, is pivotal also within the DLC film itself. Indeed, the weak adhesion of carbon coatings on metallic substrates has long been one of the main obstacles to their widespread industrial application, as it can eventually lead to spallation or delamination, making the coating entirely ineffective \cite{spallation, spallation_2, spallation_3}. Conversely, a weak adhesion with the countersurface is crucial to achieve low friction and wear \cite{Luo_2007}. For these reasons to fully exploit the potential of DLCs, it is essential to carefully design and optimize the combination of substrate, coating and countersurface and, ideally, to tune those properties that govern adhesion while growing the film itself so that it proves "sticky" on one side and unreactive on the other.\\
One of the most established strategies to improve adhesion with the substrate is to modify the substrate itself through the introduction of a metallic interlayer, which helps to relieve residual stress within the coating and provides a strong chemical bond between carbon and metal substrate atoms. This is particularly important for steel or iron substrates, where carbon diffusion during chemical vapor deposition (CVD) can lead to the formation of graphite-like layers that weaken the interface bonding \cite{C_diffusion, C_diffusion2, C_diffusion3}. Both theoretical \cite{BIBBIA, Ti_interlayer_DFT} and experimental \cite{LEPICKA201762, Cr_interlayer_WDLC, interlayer_Exp, interlayer_Exp2} studies have shown that carbide-forming metals, such as Ti, Cr and W, are highly effective as interlayers, ensuring the adhesion required for DLC to maintain its anti-wear functionality.\\
At the same time, the nature of the carbon coating itself also strongly influences its tribological properties. A higher fraction of sp$^2$ hybridized carbon atoms leads to a more graphite-like character, producing a coating with low friction coefficient, whereas a larger amount of sp$^3$ carbons leads to a diamond-like structure with superior hardness and elastic modulus \cite{60years_DLC,GRILL1999428}. Hence, it is clear that the sp$^3$/sp$^2$ ratio is a crucial factor governing adhesion and wear behavior of amorphous carbon coatings and should not be overlooked \cite{C_diffusion, lubricants10050085, YAMAMOTO20121, ma13081911}.\par
Given the numerous factors that influence the performance of carbon-based coatings, i.e., the type of substrate and interlayer material, the properties of the DLC film and those of the countersurface, experimental testing of all possible combinations just to find the best one for a specific application is extremely time and resources consuming. This is where accurate and systematic \textit{ab initio} calculations could play a role to help speed up the process through microscopic understanding and accurate datasets of theoretical predictions. Indeed, such a systematic study enables the rapid identification of optimal candidates for specific applications to be further tested experimentally, while also revealing quantitative relationships between adhesion and other fundamental physical quantities, as will be discussed in the following sections. Nevertheless, theoretical studies on diamond/metal interfaces remain limited, both in terms of the kind of metals and surface orientations investigated and in the diversity of the employed computational setup and  system modeling \cite{BIBBIA, Ti_interlayer_DFT, C100, ICHIBHA2018168, PhysRevLett.87.186103, doi:10.1021/acs.langmuir.2c02127}. This makes systematic comparisons difficult and leaves the factors governing adhesion and tribological performance poorly understood. \\
In this work, we address this gap using an \textit{ab initio} high-throughput approach, enabling a systematic evaluation and comparison of adhesion energies across technologically relevant diamond/metal interfaces. Furthermore, we propose a rigorous computational protocol for constructing the interface supercell. In contrast to the common practice of simply adapting the lattice of one component to the other, typically by straining the metallic substrate to match the diamond lattice often resulting in unphysical deformations, our approach employs the algorithm developed by Zur, which identifies the smallest common supercell between two crystalline surfaces while controlling lattice mismatch and angular distortion below user-defined limits \cite{Zur}. A detailed description of the algorithm and its implementation is provided in Section \ref{sec:methods}.\\
On the metal side, this study considers the most stable low-index terminations of Al, Ag, Au, Cr, Cu, Fe, Ir, Mg, Mo, Pt, Rh, Ti, V, W and Zn, i.e. the crystallographic orientations with the lowest surface energies, since these facets are the most likely to be exposed under equilibrium conditions. We note that chemical modifications of metal surfaces are expected to alter the adhesion energy, because the interfacial bonding is strongly affected by the local surface chemistry, as shown in previous works \cite{POLI2024160177,DAMIANI2024119555}. On the diamond side, four different low-index surfaces are chosen to represent the different types of bonds in DLC coatings: the non-reconstructed C(111)(1$\times$1) surface with dangling bond (DB) termination, its most stable reconstruction C(111)(2$\times$1) called Pandey reconstruction, the C(110)(1$\times$1) surface and the so-called Dimer reconstructed C(100)(2$\times$1) termination. From now on, we will address them as C(111), C(111)-rec, C(110) and C(100), respectively. C(111) is the only one exhibiting a sp$^3$ bond structure at the surface; while C(111)-rec, with its zig-zag surface chains and delocalized $\pi$ electrons, is representative of the sp$^2$ aromatic or graphitic layer commonly observed on the diamond/DLC surface during rubbing. C(110) also forms zig-zag $\pi$-bonded chains, but its stabilization and inertness is less pronounced than that of C(111)-rec; whereas C(100) stabilizes through surface dimers, resulting in a distorted sp$^2$-like geometry \cite{DeLaPierre, absolute_surf_energy, Bechstedt2003}.\par
The paper is organized as follows: in Section \ref{sec:methods}, we describe the computational protocol implemented in our high-throughput software for the automated construction and characterization of diamond/metal interfaces. Section \ref{sec:results} presents and discusses the results, focusing first on the trends in adhesion energies across the various diamond terminations and metal substrates and on how these relate to the intrinsic properties of the constituents, such as their surface energy. Then, the fracture behavior of the interfaces is examined to assess their mechanical stability under tensile load. Finally, electronic and structural responses upon interface formation, such as charge redistribution and modification of diamond surface reconstruction, are addressed, providing a microscopic understanding of the adhesion mechanisms.

\begin{figure}[htb!]   
 \centering
\includegraphics[width=0.4\textwidth]{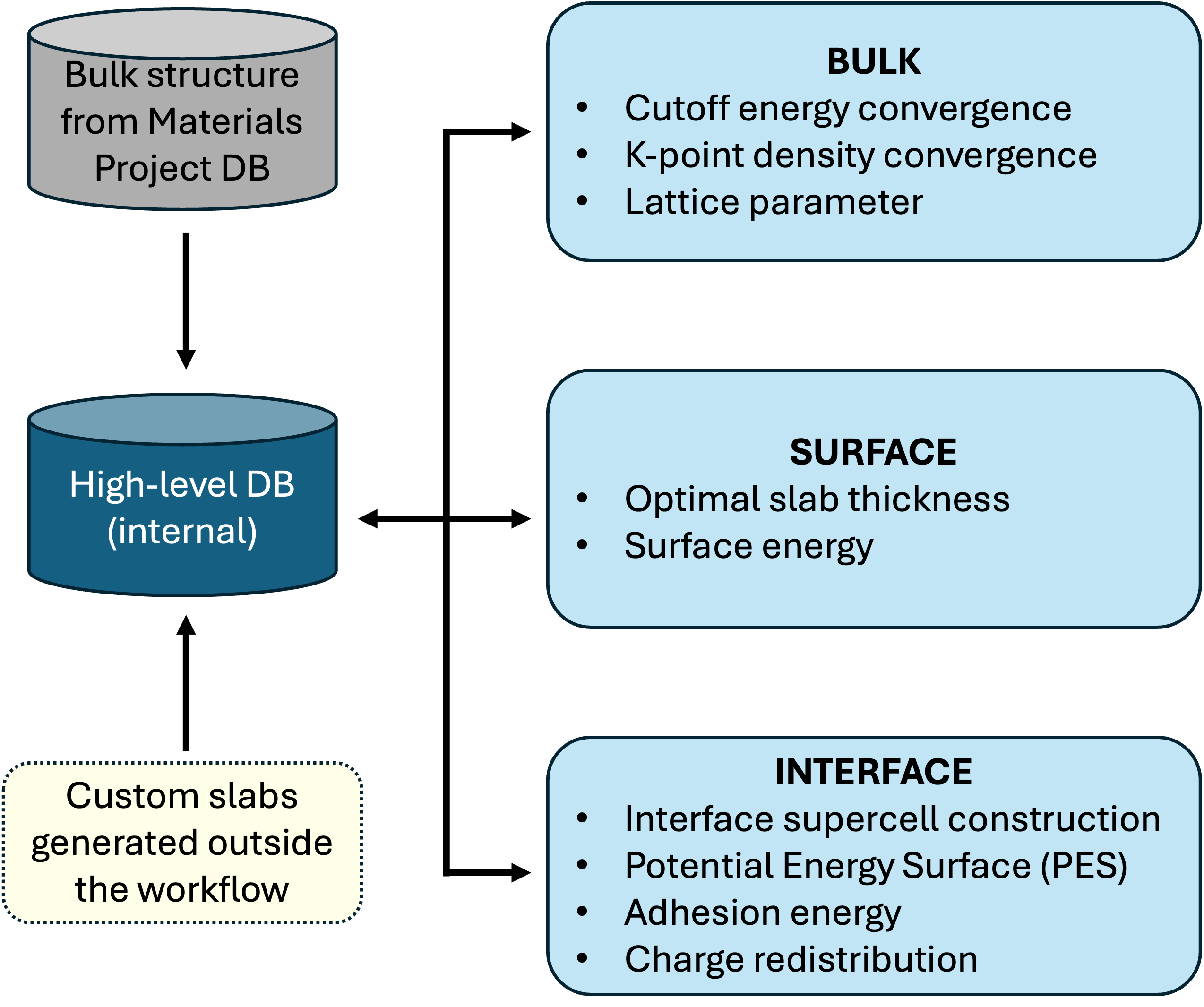}
    \caption{Schematic representation of the computational workflow implemented in TribChem to calculate solid interface properties.
   }
    \label{fig:wf}
\end{figure}

\section{METHODS} \label{sec:methods}
\begin{figure*}[htb!]   
 \centering
\includegraphics[width=1\textwidth]{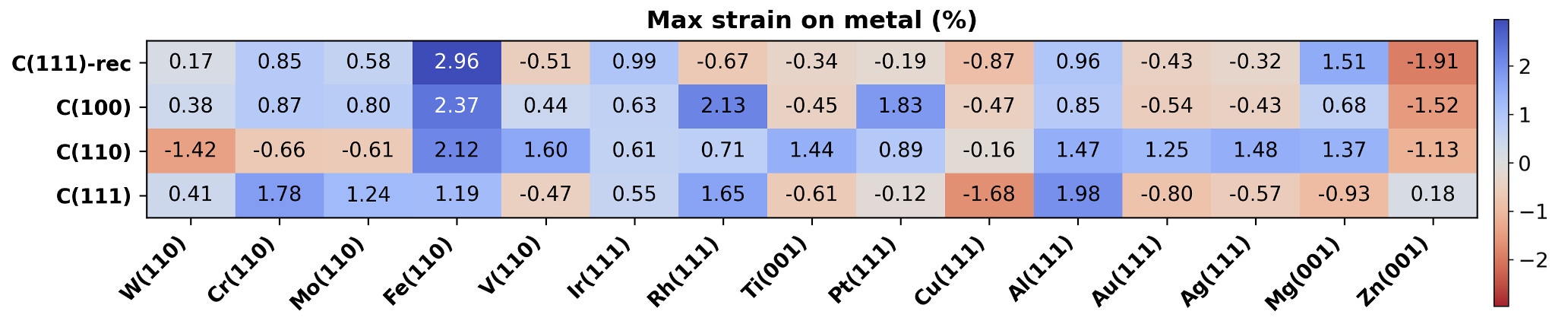}
    \caption{Maximum percentage strain applied on the metal constituent in the interface supercell. Negative values correspond to compressive strain (red regions), whereas positive values represent tensile strain (blue regions).}
    \label{fig:max_strain}
\end{figure*}

The adhesion of diamond/metal interfaces has been systematically investigated through automatized simulations using TribChem \cite{Tribchem}, an advanced software for the high-throughput atomistic study of solid-solid interfaces and of their tribological properties, developed by our group. TribChem is written entirely in Python and performs Density Functional Theory (DFT) simulations using the Vienna Ab initio Simulation Package (VASP) \cite{VASP_1,VASP_2,VASP_3}. The core of the code is the FireWorks workflow manager \cite{fireworks}, which, in addition to executing workflows, allows both the creation of custom workflows and the use of the built-in ones available in the Atomate package \cite{Atomate}. In Tribchem, most workflows are customized, but Atomate is still employed, especially to manage VASP runs and recover errors through its built-in integration with the Custodian package \cite{custodian}. Other essential libraries are Pymatgen \cite{pymatgen} and MPInterfaces \cite{mpinterfaces}, used for handling input as well as pre- and post-processing operations.
A key feature of TribChem is its systematic storing of relevant physical results. In general, the database is a fundamental component in high-throughput workflows; however, in TribChem, beyond the internal FireWorks database used for tasks management, a new high-level one is created to store physical quantities returned by the workflow in an organized manner. In particular, we use MongoDB, a NoSQL database.\\
The software has been well tested for homogeneous and heterogeneous metal-metal interfaces \cite{tribchem_homo, tribchem_hetero}, and it has been recently extended to handle the more complex structure of semiconducting materials. \par
Tribchem allows for the automatic calculation of several key parameters, including bulk modulus, surface energy, adhesion energy and charge transfer during interface formation, in order to analyze structural and tribological characteristics of these systems.\\
The workflow starts by reading the user-defined input parameters, comprising the name of the workflow to be executed, the unique identifiers of the basic bulk structures and the Miller indices that identify the slab orientation for each material. Additional parameters, such as the DFT exchange and correlation functional, whether Van der Waals corrections should be used, the maximum allowed cross-section area of the interface supercell and many others, can be supplied, otherwise default values are retrieved from a central JSON file. In this work, the Perdew-Burke-Ernzerhof generalized-gradient approximation (PBE-GGA) \cite{PBE_approx} of the exchange-correlation functional has been employed and all calculations are spin polarized. \\
A schematic representation of the computational workflow is shown in Fig.~\ref{fig:wf}. The first step consists in retrieving the input structures for the selected materials from the online Materials Project database \cite{MP_db} and storing them in our database. Both bulk structures are then optimized in parallel, by converging their lattice parameters via a fit with the Birch-Murnaghan equation of state \cite{BM} and optimizing the kinetic energy cutoff and k-point density mesh required by DFT. The higher values for energy cutoff and k-point density are chosen after all convergence loops have finished, so that the interfaces calculations will be both efficient and accurate. Subsequently, slabs are obtained by cutting the bulk along the selected directions and their thickness is converged with respect to the surface energy. The physical quantities obtained from the bulk and surface workflows for the selected metals and diamond terminations are reported in Tab.\ref{tab:bulk_surf}; while optimized computational parameters are shown in Tab.S1. The diamond surfaces characterized by complex reconstructions, namely C(111)-rec and C(100), were constructed outside the automated workflow, but still using the bulk parameters optimized by TribChem.
\begin{table}[htb!]
    \centering
\begin{tabular}{ccccc}%c|}
\hline\hline
 Material & $a$ ({\AA}) & K (GPa) & ($hkl$) & $\gamma$ (J/m$^2$) \\  
\hline
W   & 2.759 & 302 & (110) & 3.28 \\
Cr  & 2.866 & 179 & (110) & 3.19 \\
Mo  & 2.736 & 258 & (110) & 2.83 \\
Fe  & 2.452 & 179 & (110) & 2.51 \\
V   & 2.597 & 182 & (110) & 2.40 \\
Ir  & 2.738 & 348 & (111) & 2.31 \\
Rh  & 2.708 & 255 & (111) & 2.03 \\
Ti  & 2.935 & 112 & (001) & 2.02 \\
Pt  & 2.806 & 247 & (111) & 1.49 \\
Cu  & 2.569 & 140 & (111) & 1.29 \\
Al  & 2.860 &  77 & (111) & 0.84 \\
Ag  & 2.934 &  90 & (111) & 0.62 \\
Au  & 2.940 & 138 & (111) & 0.71 \\
Mg  & 3.208 &  36 & (001) & 0.54 \\
Zn  & 2.625 &  73 & (001) & 0.32 \\
\hline
\multirow{4}{*}{C} & \multirow{4}{*}{2.526} & \multirow{4}{*}{430} & (111)-rec & 3.43 \\
 & & & (100) & 4.66 \\
 & & & (110) & 5.15 \\
 & & & (111) & 5.04 \\
\hline
\hline
\end{tabular}    
    \caption{Lattice parameter ($a$), bulk modulus ($K$), Miller indexes ($hkl$) and surface energy ($\gamma$) for the considered metals and diamond structures.}
    \label{tab:bulk_surf}
\end{table}
\\As already anticipated in Section \ref{sec:introduction}, the matching of the two slabs to construct the heterointerface is performed following the Zur algorithm \cite{Zur}, implemented in the MPInterfaces library, which constructs the smallest supercell possible while satisfying specific criteria on lattice dimensions and angles strains. Further details on the algorithm are thoroughly discussed in Ref.\cite{tribchem_hetero}. The magnitude of the applied strain is scaled according to the inverse of the bulk modulus of each material, meaning that materials with higher compressibility undergo greater deformation during interface construction. However, in this case, since diamond is significantly stiffer than the metals considered, the strain was applied exclusively to the metallic component. Since lattices with different symmetries must be accommodated within a single supercell, the strain on the metal is anisotropic in some cases. Fig.\ref{fig:max_strain} reports the maximum percentage strain, while average strain values are provided in Fig.S1. For all metals, the lattice strain is below 3$\%$, which significantly reduces the impact of mechanical stress on the interface electronic properties. Details on the transformations applied to the slab cells to construct the interface supercell, as automatically computed by TribChem, can be found in the Supporting Information (SI).\\
Finally, the tribological properties of the interface such as the Potential Energy Surface (PES), adhesion energy and charge transfer at the interface are calculated. The PES maps the interaction energy as a function of the relative lateral position of the two slabs. It is therefore obtained by relaxing the slabs in the direction normal to the interface plane for all the relative lateral shifts obtained by matching the two surfaces’ high symmetry points. The resulting PES is then used to identify the minimum-energy registry. Starting from this configuration, a subsequent fully unconstrained relaxation (atomic displacements allowed in all directions) is performed. Its total energy, $E_{12}^{min}$, is the one employed to compute the adhesion energy:
\begin{equation}
    E_{adh}=\frac{1}{A}(E_{12}^{min}-E_1-E_2),
    \label{eq:adhesion}
\end{equation}
where $E_1$ and $E_2$ are the total energies of the isolated slabs, relaxed with the same settings, and $A$ is the interface supercell area. For all relaxation calculations, the electronic self-consistency threshold is $5\times 10^{-6}$ eV and the ionic relaxation is converged when residual forces are below 0.015 eV/\AA. Electronic charge redistribution upon interface formation ($\rho_{red}$) is calculated as:
\begin{equation}
    \rho_{red}=\frac{1}{2z_0}\int_{-z_0}^{z_0} \left| \frac{\Delta \rho(z)}{A} \right|dz
    \label{eq:rho_acc}
\end{equation}
where $2z_0$ is the interface distance and $\Delta \rho(z)$ is the planar average of the interface charge density difference $\Delta \rho(\textbf{r})=\rho_{12}(\textbf{r})-\rho_1(\textbf{r})-\rho_2(\textbf{r})$; $\rho_{12}(\textbf{r})$ being the volumetric charge density of the interface, while $\rho_1(\textbf{r})$ and $\rho_2(\textbf{r})$ those of the single surfaces.

\section{RESULTS AND DISCUSSION} \label{sec:results}
\begin{figure}[htb!]   
 \centering
\includegraphics[width=0.4\textwidth]{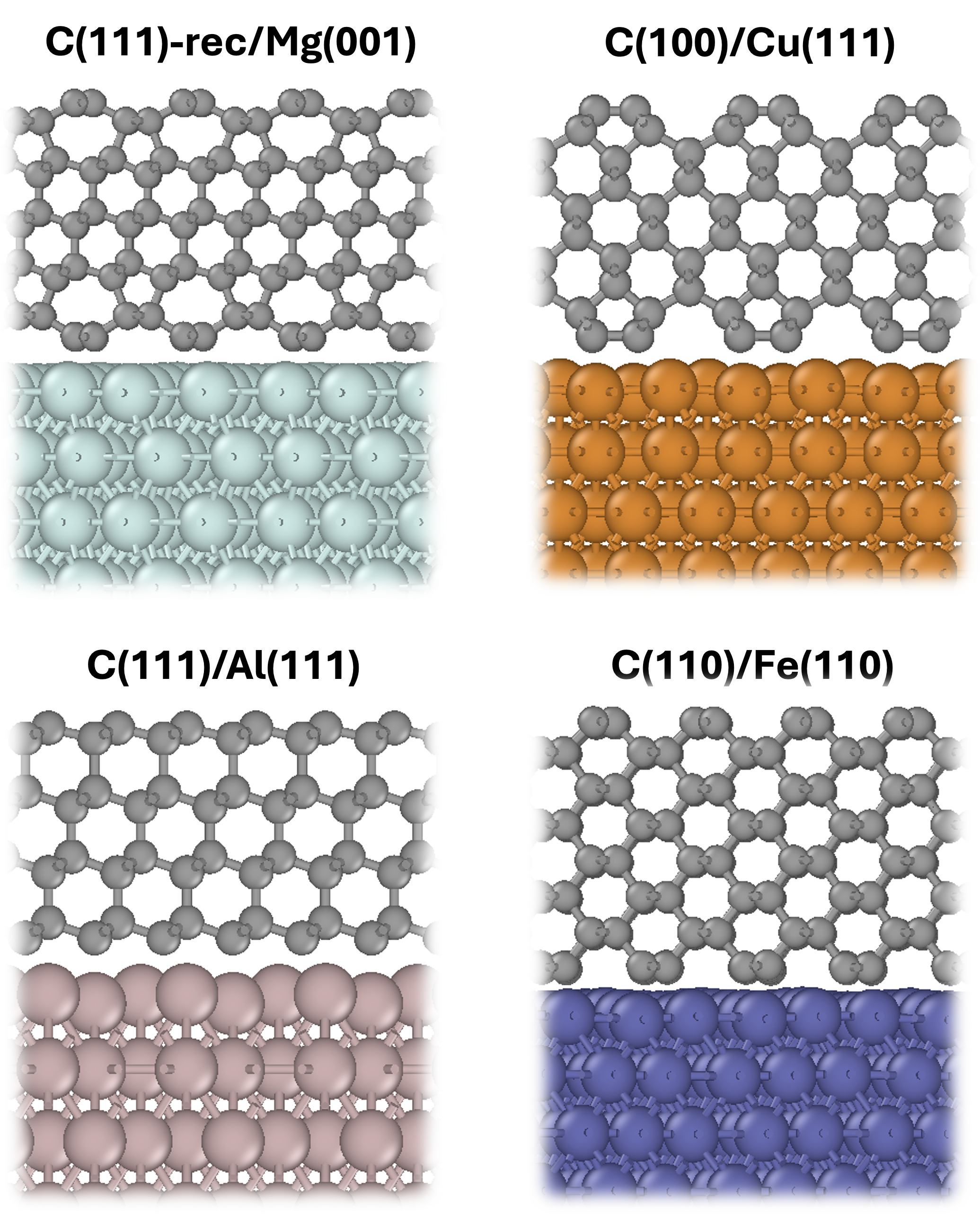}
    \caption{Atomistic models of interfaces between the four considered low-index diamond terminations and representative metallic substrates.
   }
    \label{fig:atoms}
\end{figure}

Considering a subset of technologically relevant transition metals and including Al and Mg simple metals due to their importance in alloys, the most stable facets of Ag, Al, Au, Cr, Cu, Fe, Ir, Mg, Mo, Pt, Rh, Ti, V, W and Zn have been mated to four low-index diamond surfaces: C(111), C(111)-rec, C(110) and C(100). A total of 60 diamond/metal interfaces were optimized and their adhesion were calculated. This requires to sample the whole PES and identify the energy minimum, meaning that a total of 950 DFT calculations were performed. The atomic models of interfaces formed by the four diamond terminations in contact with representative metallic substrate are shown in Fig.\ref{fig:atoms}. 
\begin{figure*}[ht!]   
 \centering
\includegraphics[width=1\textwidth]{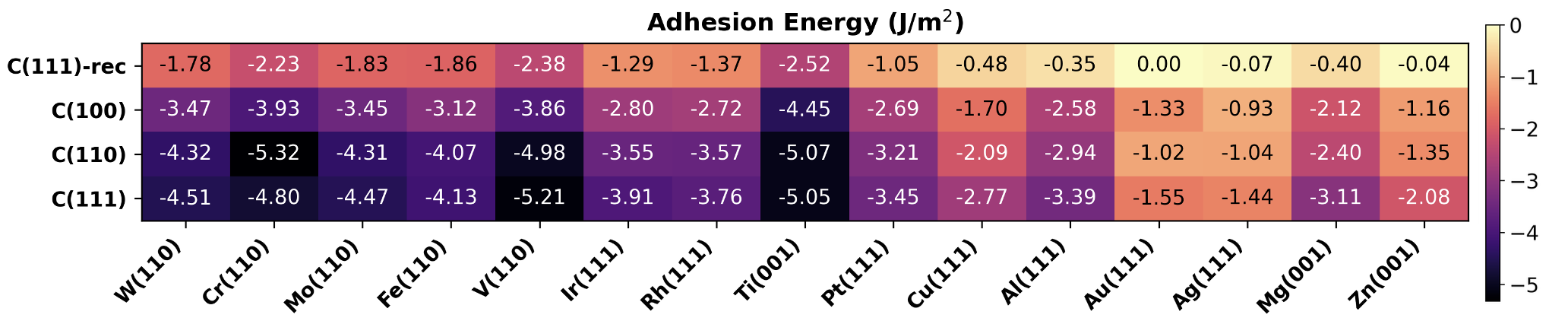}
    \caption{Adhesion energies of diamond/metal interfaces. Darker colors represents higher $|E_{adh}|$. Metals are ordered from left to right according to decreasing surface energy.
   }
    \label{fig:adhesion}
\end{figure*}
\subsection{Adhesion energy}
Adhesion energies, as calculated from Eq.\ref{eq:adhesion}, are reported in Fig.\ref{fig:adhesion}. In general, constituents with higher surface energies form an interface characterized by higher adhesion (lower left region of Fig.\ref{fig:adhesion}). This comprises in particular C(110) and C(111) diamond terminations in contact with reactive transition metal surfaces, like the bcc(110) early transition metals W, Cr, Mo, V. By contrast, all diamond terminations adhere less to metals which possess completely filled d-shells, such as Cu, Ag, Au.
%This suggests that metals with partially filled d-shells, e.g. W, Cr, Ti, represent suitable candidates for use as interlayers to boost adhesion between the DLC coating and the substrate, as already demonstrated in the context of 2D material coatings to improve interfacial stability [REF]. 
When examining the values in Fig.\ref{fig:adhesion} column by column, i.e. at fixed metal substrate, diamond terminations with higher surface energy exhibit stronger adhesion across all metals. An interesting comparison can be made between C(111), which exposes surface sp$^3$ carbons, and C(111)-rec, characterized by chains of sp$^2$ hybridized carbon atoms. While previous work reported adhesion reduction at C/Cu due to surface graphitization \cite{DAMIANI2024119555}, here we see that this behavior is general and that surface graphitization drastically reduces adhesion across all metals. This suggests that while growing the DLC film, for an effective coating that maximizes the adhesion to the substrate while reducing adhesion and friction with any countersurface, a gradient of the sp$^2$/sp$^3$ fraction should be built with larger sp$^2$/sp$^3$ ratio. Conversely, when examining Fig.\ref{fig:adhesion} row by row, i.e. by varying the metal substrate while keeping the diamond termination fixed, it is possible to observe that metals with a partially occupied d-shell adhere most strongly, especially Ti, Cr and V. In particular, Ti shows strong adhesion to all diamond terminations, despite its relatively low reactivity, quantified by its surface energy. This behaviour can be attributed to its pronounced carbide-forming ability compared to other metals (carbide formation enthalpy of TiC per Ti atom $\Delta H_f$=–185.2 kJ/mol \cite{BIBBIA} and TiC formation energy of -0.808 eV/atom \cite{MP_db}).\\
%In DLC coatings used to minimize wear and friction, it is therefore ideal to control the exposure of sp$^3$ and sp$^2$ terminations at the surface: during deposition, if sp$^3$ terminations are in contact with the substrate, the coating will adhere strongly, preventing spallation. Conversely, if sp$^2$ bonds are exposed, the coating will be more effective in reducing friction and wear against a countersurface.\\
These results validate our computational protocol, as it correctly predicts that carbide-forming metals such as Cr, Ti, W and V exhibit the strongest adhesion to diamond coatings, consistent with their widespread use as interlayers between DLC and metallic substrates. %Moreover, such a systematic study enables the rapid identification of optimal candidates for specific applications to be further tested experimentally, while also revealing quantitative relationships between adhesion and other fundamental physical quantities, as will be discussed in the following sections.

\subsection{Adhesion energies from intrinsic constituent
properties}
\begin{figure}[htb!]   
 \centering
\includegraphics[width=0.48\textwidth]{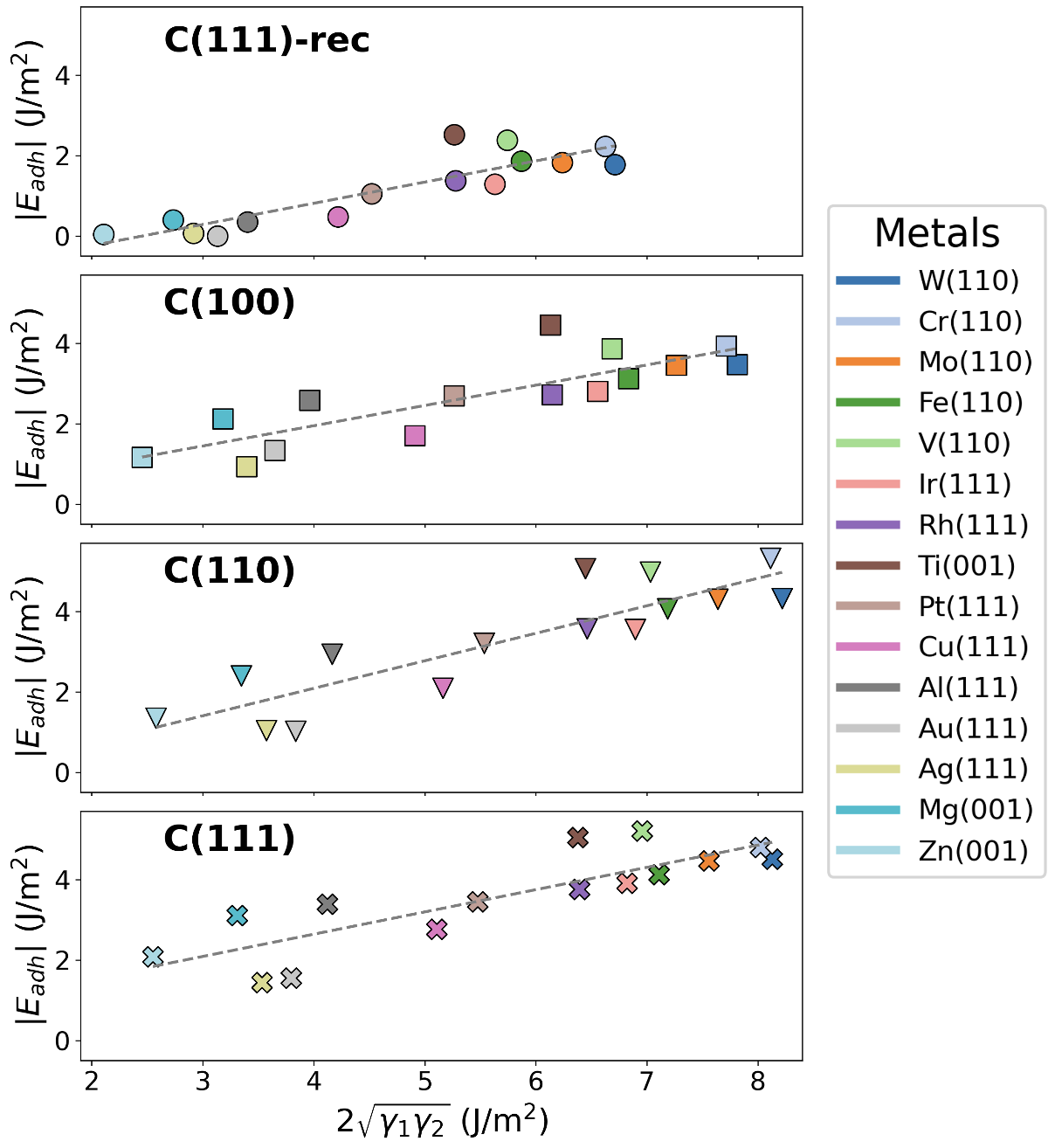}
    \caption{Adhesion energies as a function of the geometric mean between diamond and metal surface energies. Metal substrates are differentiated by colors, while symbols represent different diamond terminations. The grey dashed line represent the linear fit. The corresponding fitting parameters and quality indicators are summarized in Tab.\ref{tab:linear_fit}. Data are plotted using the same scale. 
   }
    \label{fig:adhesion_vs_surfene_single}
\end{figure}

To reduce the computational cost of simulating the full interface, it would be useful to establish an empirical relationship linking the adhesion energy to the intrinsic properties of the individual components. For hetergoeneous metallic interfaces, it was demonstrated that adhesion energy can be approximated by the geometric mean of the surface energies of the single constituents: $E_{adh}=-2\sqrt{\gamma_1\gamma_2}$ \cite{tribchem_hetero}. This expression originates from simplified physical models, where the adhesion energy is defined as $E_{adh}=\gamma_{12}-\gamma_1-\gamma_2$; $\gamma_1$ and $\gamma_2$ being the surface energies of the two constituents and $\gamma_{12}$ the interfacial energy, i.e., the cost of creating a unit area of interface starting from the corresponding bulk of the two materials. As reported in Ref.\cite{Israelachvili2011}, the latter can be expressed as: $\gamma_{12}=\gamma_1+\gamma_2-2\sqrt{\gamma_1\gamma_2}$, thereby obtaining the relation given above. \\
The effectiveness of this relationship is further confirmed for diamond/metal interfaces, as illustrated in Fig.\ref{fig:adhesion_vs_surfene_single}, where $|E_{adh}|$ is plotted against $2\sqrt{\gamma_1\gamma_2}$ at fixed diamond termination to account for the different nature of surface bonds. $R^2$ and RMSE values along with the linear fit parameters are reported in Tab.\ref{tab:linear_fit}. In all cases, the relatively high $R^2$ values, lying in the range 70-79$\%$, indicate that the overall trend predicted by the model is robust; while deviations in slope and intercept from the ideal values of 1 and 0, respectively, highlight that predicting adhesion energy is a complex task, governed by the interplay of multiple factors, such as chemical and structural variations at the interface. The correlation can be further improved by adopting a more advanced data driven strategy, such as the Sure Independence Screening and Sparsifying Operator (SISSO) algorithm \cite{SISSO}. This method has already been successfully applied to heterogeneous metallic interfaces to identify the best descriptors and the optimal mathematical expression fitting the training data, by combining multiple material properties like cohesive energies and bulk moduli \cite{tribchem_hetero}. Here, however, our goal is to provide a physical and easily interpretable message. For this reason, we focus on the term that carried the largest weight in the previous SISSO-derived expression,  i.e. $\sqrt{\gamma_1\gamma_2}$, and that also has the most immediate physical meaning and experimental accessibility, rather than presenting a more complex multi-parameter expression.

\begin{table}[htb!]
    \centering
\begin{tabular}{ccccc}%c|}
\hline\hline
           & Slope & Intercept (J/m$^2$) & $R^2$ & RMSE \\  
\hline
C(111)-rec & 0.53 & -1.30& 79$\%$ & 40$\%$ \\
C(100)  & 0.50 & -0.07 & 70$\%$ & 57$\%$ \\
C(110)  & 0.68 & -0.65 & 77$\%$ & 67$\%$ \\
C(111)  & 0.55 & 0.43 & 72$\%$ & 62$\%$ \\
\hline
\hline
\end{tabular}    
    \caption{Linear fit parameters, $R^2$ values and root mean square error (RMSE) for the correlation between adhesion energy $|E_\mathrm{adh}|$ and the geometric mean of surface energies $2\sqrt{\gamma_1\gamma_2}$ for diamond/metal interfaces, calculated separately for each diamond termination.}
    \label{tab:linear_fit}
\end{table}
%The effectiveness of this relationship is further confirmed for diamond/metal interfaces, as illustrated in Fig.\ref{fig:adhesion_vs_surfene_single}, where $|E_{adh}|$ is plotted against $2\sqrt(\gamma_1\gamma_2)$ at fixed diamond termination to account for the different nature of surface bonds. In all cases, a strong correlation is observed, with $R^2$ values in the range 69–78$\%$, as reported in Tab.~\ref{tab:linear_fit} along with RMSE values and linear fit parameters. The slopes, which range from 0.5 to 0.7, are consistently below the ideal value of 1, while the nonzero intercepts indicate systematic deviations that depend on the surface termination, reflecting the influence of surface bonding on adhesion strength.

\subsection{Fracture location}
\begin{figure}[ht!]   
 \centering
\includegraphics[width=0.48\textwidth]{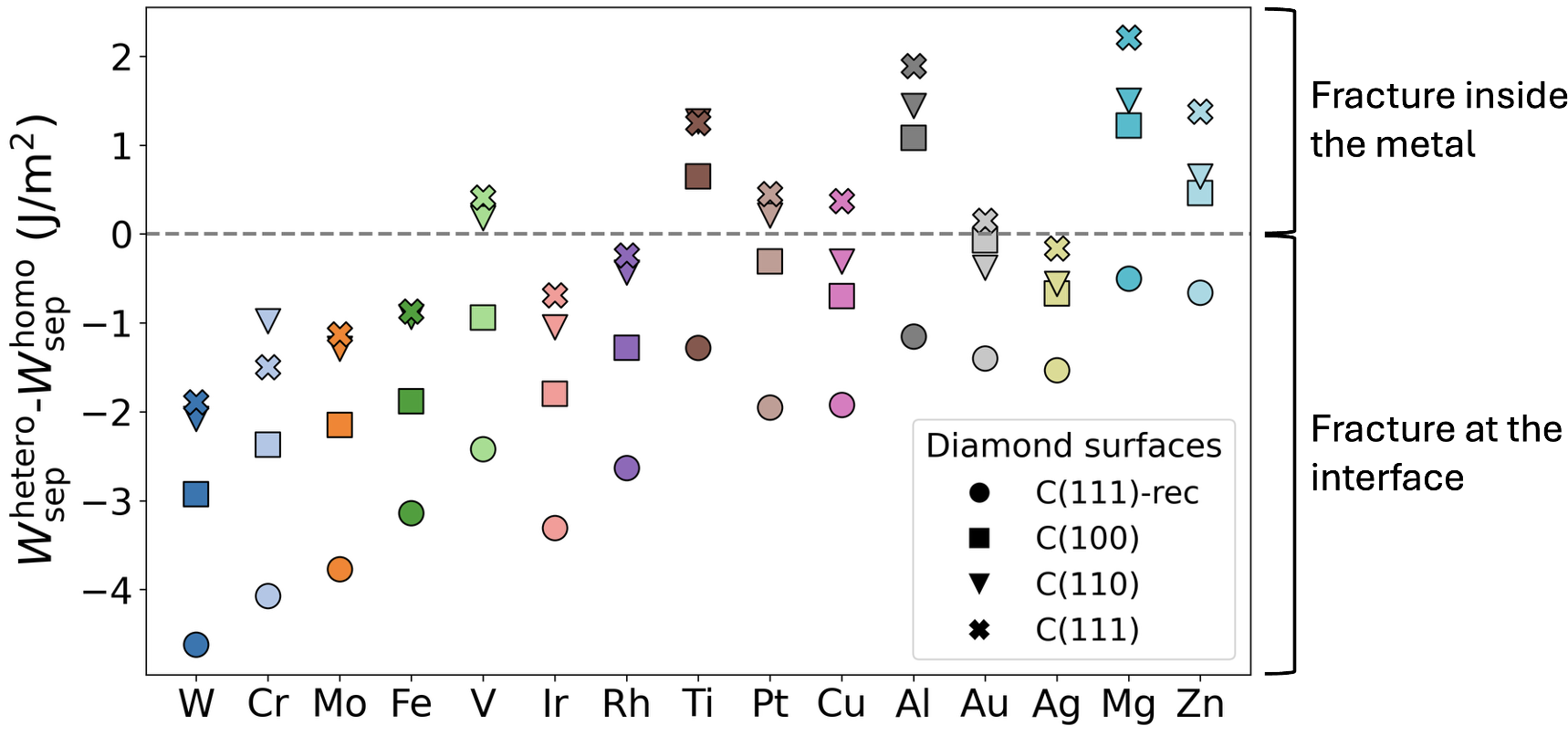}
    \caption{Difference between work of separation, $W_{sep}=-E_{adh}$, of the  heterogeneous diamond/metal interface and $W_{sep}$ of the corresponding homogeneous metallic interface from Ref.\cite{tribchem_homo}. Metallic species are differentiated by colors, while symbols represent different diamond terminations. If the difference is higher (lower) than 0 the fracture is likely to happen inside the metal (at the interface).  
   }
    \label{fig:fracture_location}
\end{figure}

In the context of wear, adhesion energy of an interface plays a key role in determining coating durability. Whether fracture under tensile load initiates within the bulk metal or at the metal/diamond interface directly affects the resistance of the coating to delamination and, consequently, to wear. A practical way to assess this is by comparing the work of separation, $W_{sep}^{hetero}=-E_{adh}$, of the diamond/metal interface with the interlayer cohesive energy of the corresponding bulk metal, a property that can be quantified by the work of separation of the homogeneous metallic interface $W_{sep}^{homo}$. If the latter is lower than the former, fracture is expected to occur inside the metal; otherwise, the interface itself becomes the weakest link. \\
Fig.\ref{fig:fracture_location} shows the difference between the work of separation of the semiconductor/metal interface and interlayer cohesive energy of the metal. When C(111)-rec is involved, the fracture always occurs at the interface, reflecting the fact that surface graphitization of diamond leads to extremely weak bonds with the metal and, consequently, to very low adhesion, increasing the risk of delamination. For metals with low surface energy ($\sim W_{sep}^{homo}$/2), e.g., Mg, Zn, Al, fracture may occur within the metal itself, especially when in contact with the highly reactive C(111) surface. On the other hand, metals with partially filled d-shells (W, Cr, Mo, Fe) exhibit extremely high cohesion energies, which prevent failure from taking place within the metal phase.

\subsection{Structural and electronic modifications of diamond reconstructions upon adhesion to metals}
\begin{figure*}[htb!] 
 \centering
  \subfloat[]{\includegraphics[width=0.48\textwidth]{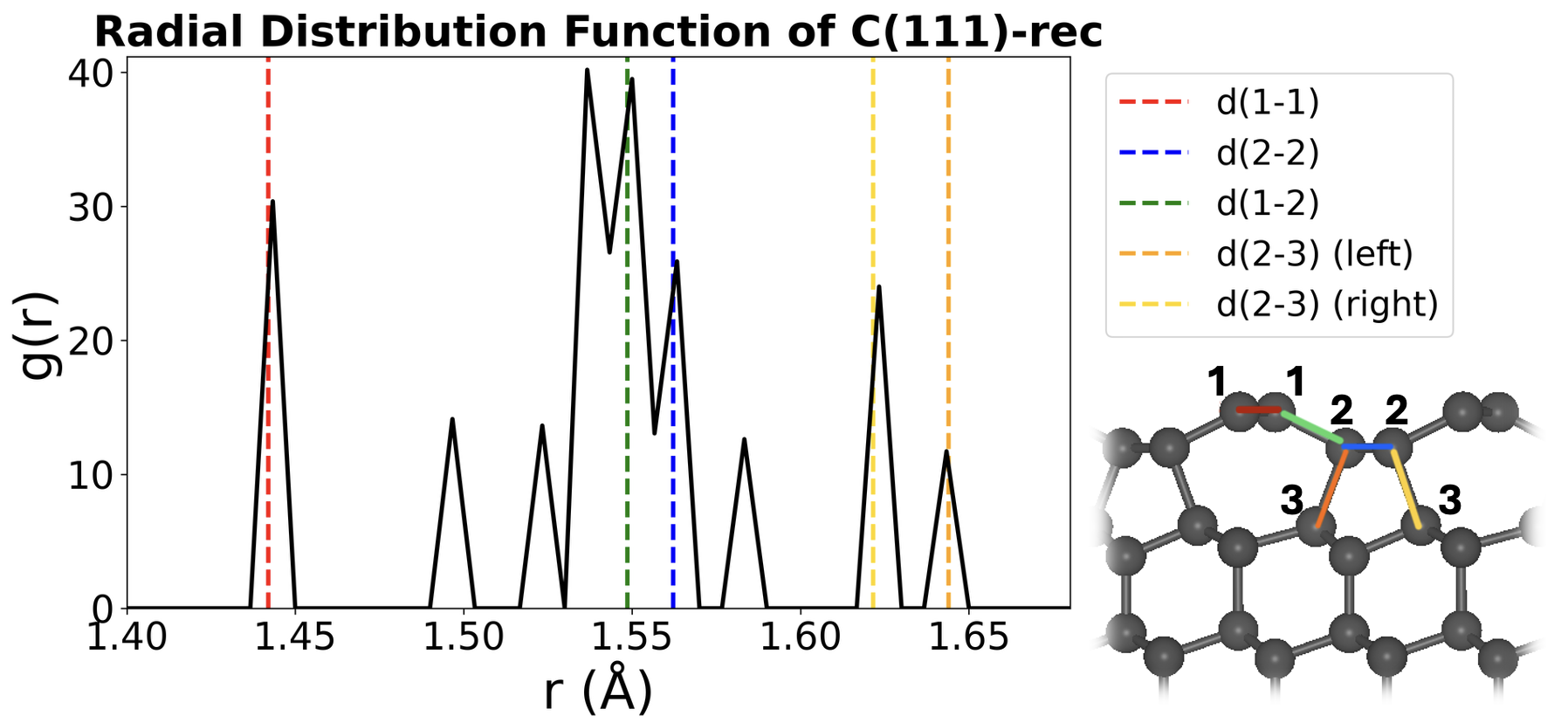}}\\
  \subfloat[]
{\includegraphics[width=1\textwidth]{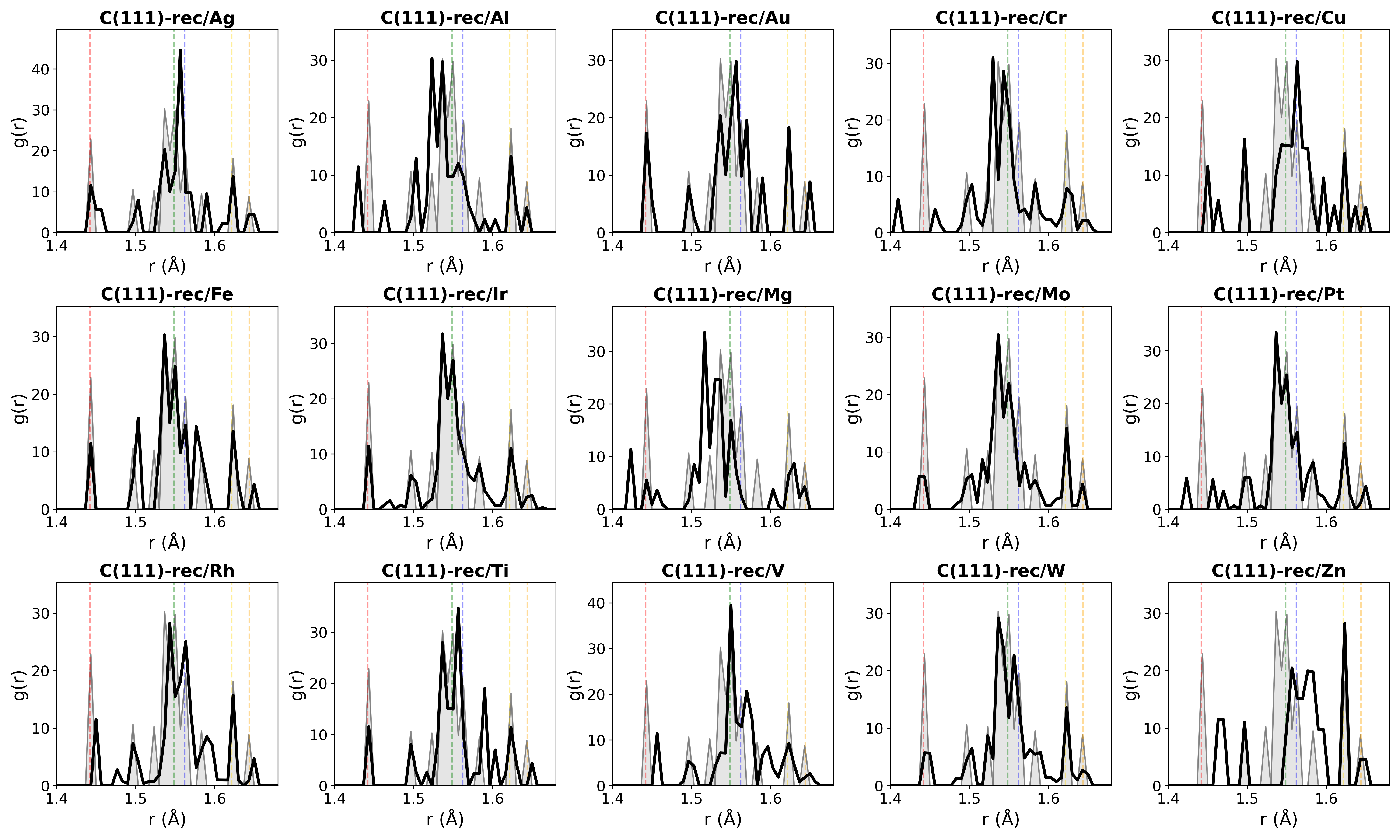}}\\
    \subfloat[]
    {\includegraphics[width=0.7\textwidth]{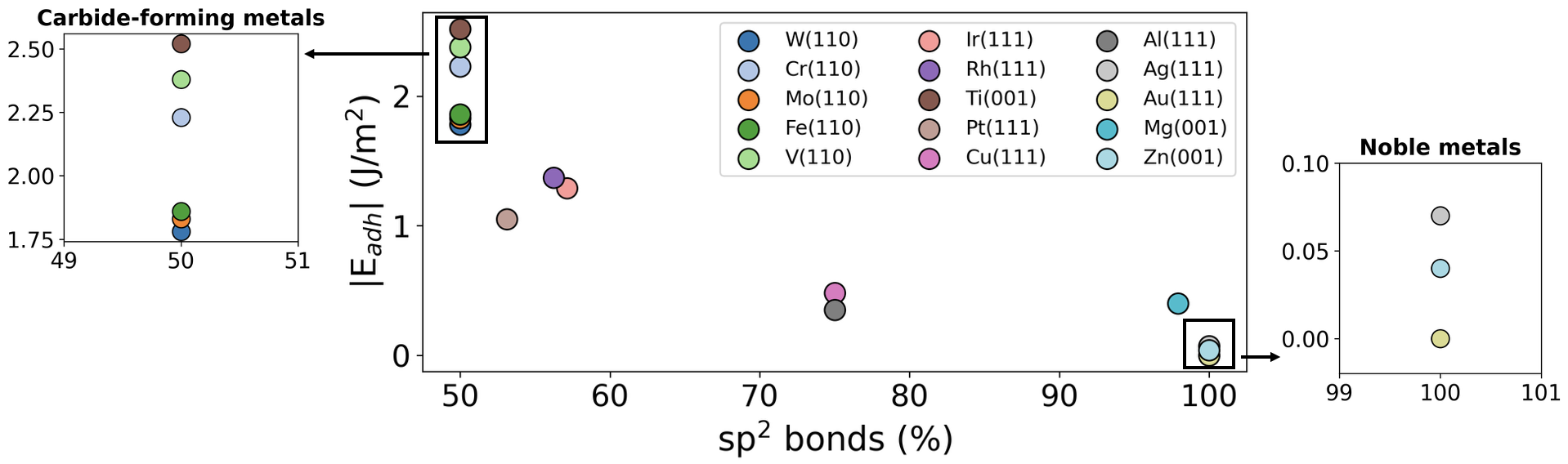}}
    \caption{(a) Radial distribution function of pristine C(111)-rec. Coloured dashed lines identify peaks corresponding to the interatomic distances highlighted in the adjacent schematic. (b) Radial distribution function of C(111)-rec in the optimized interface geometry (black line), compared with the $g(r)$ of the pristine surface (plotted in the background with faded lines). (c) Percentage fraction of sp$^2$ bonds preserved by the C(111)-rec surface after its interaction with metal substrates. Insets highlight the two densely populated regions corresponding to carbide-forming metals (50$\%$ sp$^2$ preservation) and noble metals (100$\%$ sp$^2$ preservation).
   }
    \label{fig:gr_multiplot}
\end{figure*}
To gain insight into the structural modifications of diamond reconstructions upon interface formation, we analyzed the radial distribution function $g(r)$ of the carbon atoms. The $g(r)$ describes the weighted probability of finding a pair of atoms at a given distance $r$, providing direct information on the local bonding environment \cite{rdf_1,rdf_2,rdf_3}. This method is widely exploited in molecular dynamics simulations to investigate short- and medium-range order in both crystalline and amorphous systems, as it offers a powerful way to capture structural rearrangements \cite{MD_1,Md_3}. In the present case, monitoring changes in the $g(r)$ permits to detect how the reconstruction of the diamond surface evolves upon contact with the metallic counterpart, a key aspect since geometric rearrangements are known to influence the surface electronic structure \cite{dott_marsili} and therefore its adhesion strength. \\
The $g(r)$ of the standalone C(111)-rec surface is shown in Fig.\ref{fig:gr_multiplot}(a). Peaks corresponding to the interatomic distances within the surface atoms are indicated by vertical dashed lines and illustrated in the adjacent schematic. Remaining peaks correspond to the bulk diamond structure. Here the focus is on the peak at 1.44 {\AA}, which corresponds to the distances between atoms in the surface zig-zag chain. The area under a peak is directly related to the average number of atoms in the interval [$r_1$, $r_2$] surrounding a given atom, i.e., the number of bonds per atom contributing to that peak, namely \cite{CN}:
\begin{equation}
    N=4\pi n\int_{r_1}^{r_2}g(r)r^2dr
\end{equation}
where $n$ is the average atomic density (total number of atoms over cell volume). For C(111)-rec, the integral of the peak at 1.44 Å therefore provides a measure of the number of surface sp$^2$ bonds per atom. \\
To assess the impact of metal contact on the surface graphitization, Fig.\ref{fig:gr_multiplot}(b) shows a qualitative comparison of the $g(r)$ of the pristine C(111)-rec surface (plotted in the background with faded lines) with that of the same surface but within the interface geometries (black line). Different types of structural modifications are observed depending on the metal: in some cases, the peak intensity decreases without significant shift or distortion (Mo, W, Ti, Fe, Ir), while in others, dimerization (Cr, Al, Cu, Pt, Mg) or a shift in $r$ (Rh, V, Zn) is evident, reflecting a more pronounced perturbation of the surface reconstruction. Quantitatively, integrating the peak provides a measure of how many sp$^2$ bonds remain after interaction with each metal. For example, the weak interaction with Au leaves the peak essentially unaltered, showing only minor broadening and indicating that all sp$^2$ bonds are preserved. In contrast, Cr, which exhibits strong adhesion with C(111)-rec, induces partial dimerization of the chain and reduces the number of sp$^2$ bonds by roughly 50$\%$. The adhesion energy of all C(111)-rec/metal interfaces as a function of the fraction of sp$^2$ bonds preserved relative to the pristine surface is plotted in Fig.\ref{fig:gr_multiplot}(c). Two characteristic regimes emerge, highlighted by the insets: weakly interacting metals, i.e., those with completely filled outer electron shells (Ag, Au, Zn), preserve essentially the original number of sp$^2$ bonds of the C(111)-rec surface, consistently with their weak adhesion; conversely, more reactive metals are effectively able to perturb the surface sp$^2$ network, leading to stronger adhesion. In particular, carbon-reactive/carbide-forming metals (Ti, Cr, Mo, W, V, Fe) preserve only about half of the original sp$^2$ bonds. This behavior is highly relevant for coating applications, where metallic interlayers are commonly introduced to improve the bonding between DLC films and their substrates. In DLC systems with a high density of surface sp$^2$ bonds, the interaction with the metal counteracts graphitization, responsible for the weak adhesion to the substrate, by promoting the rehybridization of surface carbon atoms.\\
The $g(r)$ analysis was also performed for other diamond terminations exhibiting surface sp$^2$ bonds, namely C(100) and C(110) (see Fig.S2 and S3). In general, interaction with the metallic surfaces does not lead to major perturbations of the $g(r)$ peak. For C(100), the main effect is the reduction in peak intensity, accompanied by minor shifts in the case of Rh, V and Ir. Similarly, C(110) generally exhibits only a decrease in peak intensity, with more noticeable effects observed for Zn, showing a significant peak shift, and Mg, where dimerization of the surface chains occurs. Quantitatively, both terminations experience an overall reduction of around 50$\%$ of the original surface sp$^2$ bonds, with the exception of C(110)/Au and C(110)/Ag, where $\sim$54$\%$ and 59$\%$ of the original bonds are preserved, respectively. \\
Although no change in reconstruction is expected for the unreconstructed C(111) termination upon contact (its primary response being the saturation of its highly reactive dangling bonds through C–metal bond formation), we extended the $g(r)$ analysis to C(111) and to the metallic counterparts as a general tool to track slab structural evolution upon interface formation (see SI). In C(111) the surface bilayer sp$^3$ C–C peak is mostly only reduced/broadened, with the strongest perturbations for Cr, Ti, V; for the metals, the largest peak changes are observed for the more strongly adhering transition metals (e.g., V, Ti, Mo, Cr).\par
\begin{figure}[htb!]   
 \centering
\includegraphics[width=0.48\textwidth]{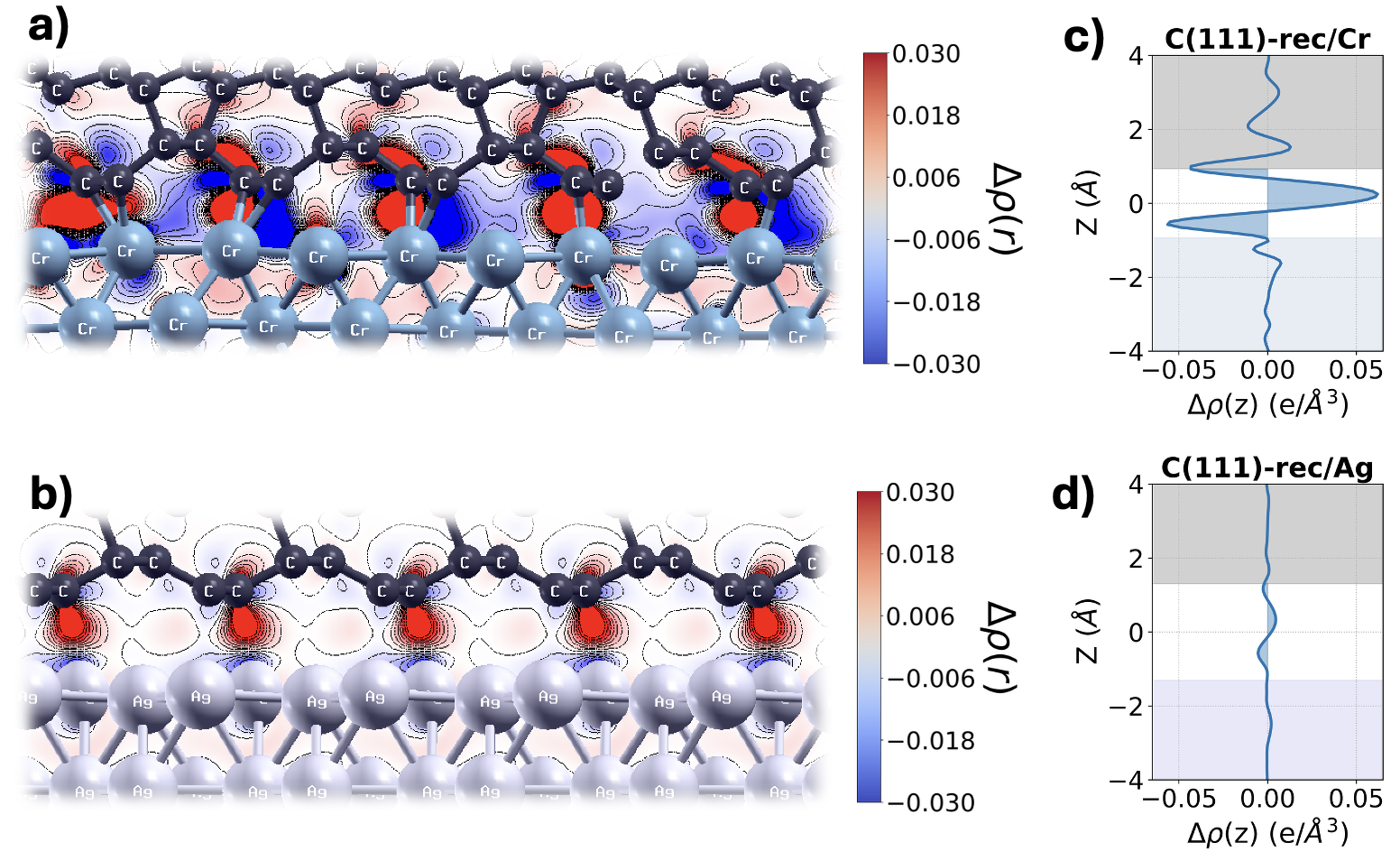}
    \caption{Volumetric charge density difference, $\Delta \rho(\mathbf{r})$, for C(111)-rec/Cr (a) and C(111)-rec/Ag (b): electron accumulation and depletion are shown in red and blue, respectively, with a colorbar indicating the values. Line profiles along the $z$ direction are reported in panels (c) for Cr and (d) for Ag. The gray-shaded regions indicate the position of the diamond slab, while the light blue and violet shaded regions mark the positions of the Cr and Ag slabs, respectively. The area highlighted under the curve corresponds to $\rho_{\text{red}}$, the descriptor calculated with Eq.\ref{eq:rho_acc}.
   }
    \label{fig:charge_dis}
\end{figure}

  %The stronger perturbation of the sp$^2$ bonds of C(111)-rec upon contact with metals highlights their superior stability and robustness relative to those of C(100) and C(110); a characteristic that is also reflected in the fact that C(111)-rec exhibits the lowest surface energy among all the considered terminations (see Tab.\ref{tab:bulk_surf}).\par
The rehybridization of the C(111)-rec surface chains is clearly reflected in the redistribution of electronic charge upon contact with the metal. As examples, we consider the interfaces with Cr and Ag, representing two extreme cases: the Cr interface exhibits dimerization of the surface chains and strong adhesion ($E_{adh}=-2.23$ J/m$^2$), while the Ag interface shows only a slight broadening of the peak, corresponding to weak adhesion ($E_{adh}=-0.07$ J/m$^2$). The volumetric charge difference, $\Delta \rho(\mathbf{r})$, shown in Fig.\ref{fig:charge_dis}(a) and (b), reveals that the charge rearrangement at the Cr interface is significantly more complex than at Ag. This is further quantified by the line profiles of $\Delta \rho(\mathbf{r})$ along the $z$ direction, reported in Fig.\ref{fig:charge_dis}(c) and (d). In both cases the metal just accumulates charge in the interfacial region, while the surface carbon atoms reorganize their electronic charge density. Importantly, for the interface with Cr, this reorganization is far more pronounced: charge is removed at the surface carbon atom atomic plane where the C-C double bonds of the Pandey chain laid, moving towards the interface region, in line with the observed dimerization of the surface chains.
%Since modifications of the surface geometry directly affect its electronic properties, these structural rearrangements are expected to have a significant impact on the electronic properties of the interface, and therefore on its adhesion strength.

\subsection{Charge density redistribution at the interface} \label{charge_acc}
\begin{figure}[htb!]   
 \centering
\includegraphics[width=0.48\textwidth]{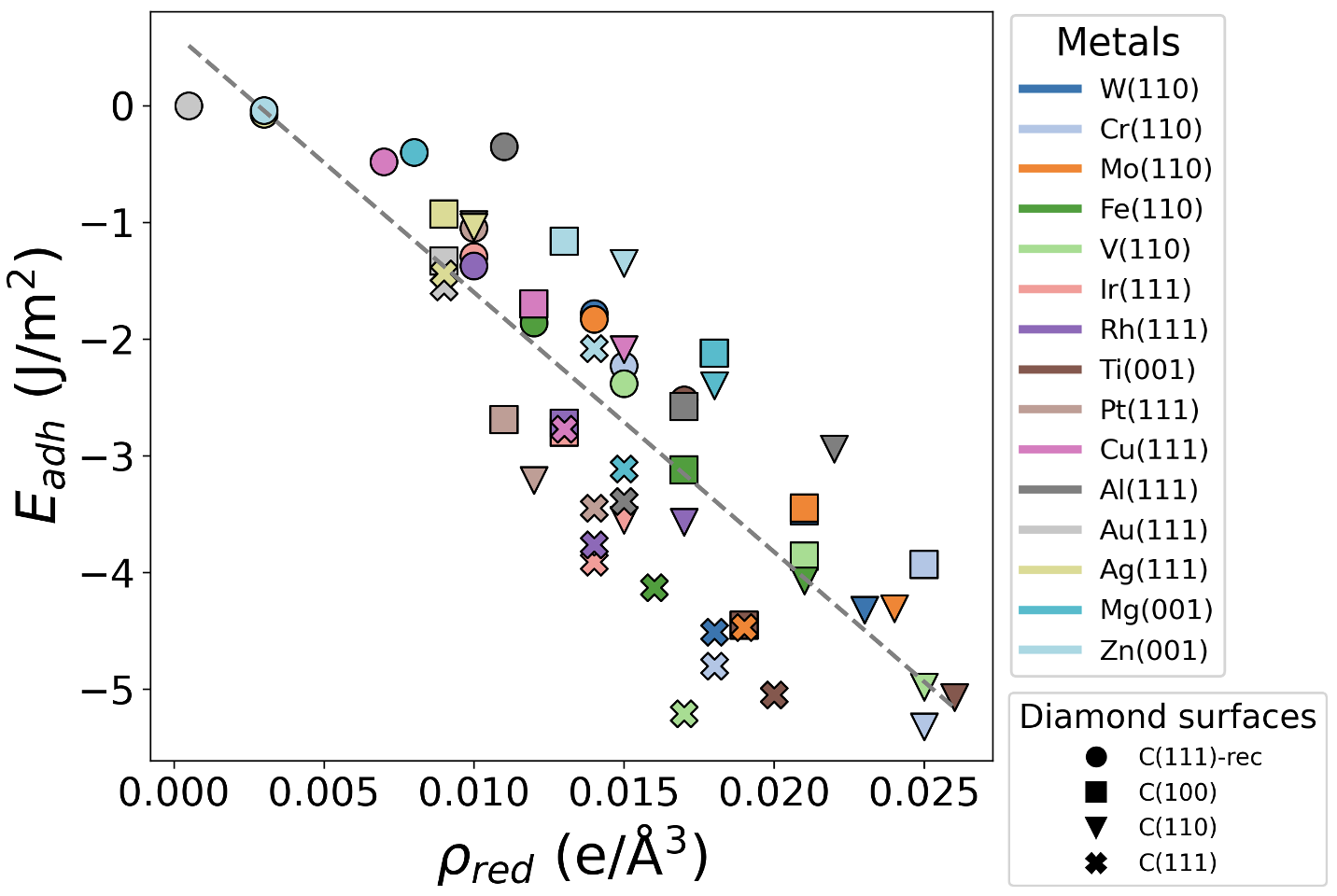}
    \caption{Adhesion energies as a function of the charge redistribution upon interface formation. Metal substrates are differentiated by colors, while symbols represent different diamond terminations. The grey dashed line represent the linear fit, with $R^2=71\%$ and RMSE=$79\%$.
   }
    \label{fig:adhesion_vs_charge}
\end{figure}

The structural modifications of the diamond reconstructions directly affect the electronic charge redistribution at the interface. Previous studies have shown that the redistribution of electronic charge upon contact, quantified by the descriptor $\rho_{\text{red}}$ (Eq.~\ref{eq:rho_acc}), correlates well with adhesion across a wide variety of homogeneous interfaces, from weakly interacting van der Waals systems (e.g., graphene) to covalently bonded materials such as diamond \cite{rho_red}, as well as heterogeneous metallic junctions \cite{tribchem_hetero}.\\
In the present case, the semiconductor-metal interaction introduces additional complexity, as surface carbon chains can undergo rehybridization and other structural rearrangements, modifying the local electronic environment. These factors, together with the broader variability of bond character, naturally increase the scatter. Nevertheless, a good correlation between adhesion energy and $\rho_{\text{red}}$ is still present, with $R^2=71\%$ (Fig.~\ref{fig:adhesion_vs_charge}), supporting $\rho_{\text{red}}$ as a reliable predictor also for this class of interfaces. Notably, even for homogeneous interfaces the correlation is not perfect: in the reference study the metallic class (the most populated subset) corresponds to $R^2<0.8$ within a linear model, reflecting the sensitivity of $\rho_{\text{red}}$ to the detailed bonding chemistry at the interface \cite{rho_red}.\par
\begin{figure}[htb!]   
 \centering
\includegraphics[width=0.48\textwidth]{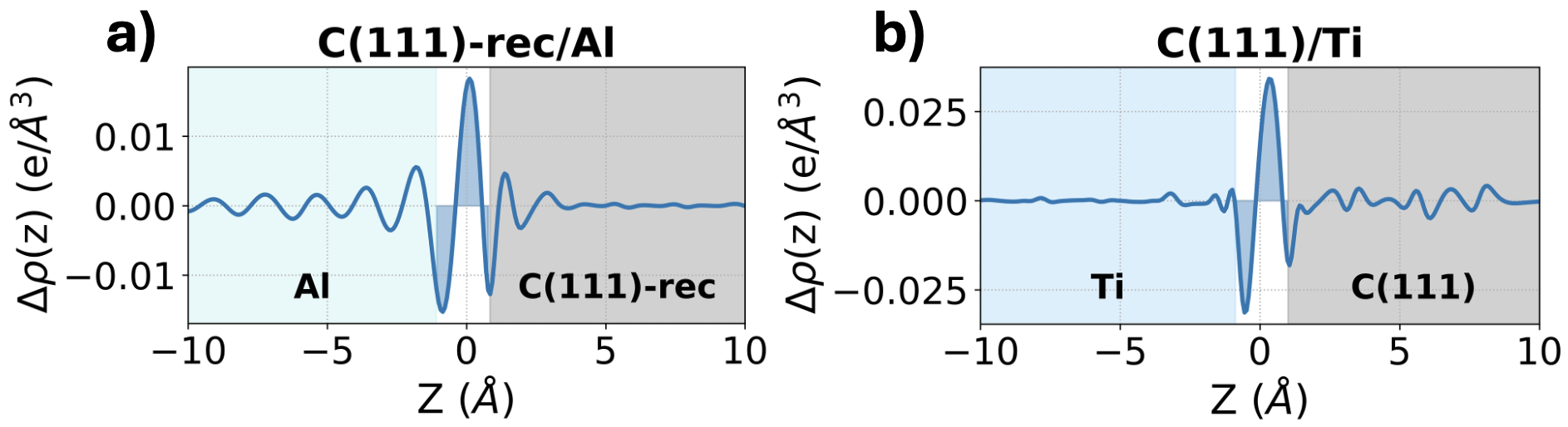}
    \caption{Line profiles along the $z$ direction of $\Delta \rho(\mathbf{r})$ are reported in panel (a) for C(111)-rec/Al and (b) for C(111)/Ti. The gray-shaded regions indicate the position of the diamond slab, while the light blue regions mark the positions of the Al and Ti slabs. The area highlighted in blue under the curve corresponds to $\rho_{\rm red}$, the descriptor calculated with Eq.\ref{eq:rho_acc}.
   }
    \label{fig:outliers}
\end{figure}

To gain insight into the origin of the residual scatter, here we analyze two opposite outliers: C(111)-rec/Al exhibits an anomalously large $\rho_{red}$ relative to its modest adhesion, whereas C(111)/Ti shows comparatively low $\rho_{red}$ despite high adhesion. To rationalize these systems, we inspected the planar averaged charge displacement $\Delta\rho(z)$, which reveals that the spatial extent of the electronic response differs markedly between the two interfaces. In C(111)-rec/Al (Fig.\ref{fig:outliers}(a)), $\Delta\rho(z)$ is not confined to the immediate contact region but extends deep into the Al slab. This is consistent with metallic screening: delocalized $sp$ electrons redistribute over several atomic layers, forming an interface dipole and yielding an oscillatory $\Delta\rho(z)$ on the Al side. This screening can lead to a large $\rho_{red}$ (defined as the integral of $|\Delta\rho(z)|$ within the interfacial window) without translating into a proportional adhesion gain, because it is less directly tied to the formation of short, directional interfacial bonds. Consistently, the average interfacial separation is relatively large (C–Al = 2.14 {\AA}), and the Al slab shows negligible first interlayer relaxation ($d_{12}$: from 2.334 to 2.336 {\AA}), both indicative of limited chemisorption-driven accommodation. \\
In contrast, for C(111)/Ti (Fig.\ref{fig:outliers}(b)) the $\Delta\rho(z)$ perturbation extends into the diamond slab, indicating that the interaction modifies the carbon electronic structure beyond the topmost layer. This behavior is compatible with strong chemisorption (consistent with C 2$p$–Ti 3$d$ hybridization) that enhances adhesion, and it is supported by the shorter interfacial distance (C–Ti = 1.95 {\AA}) and by a clear substrate relaxation on the Ti side ($d_{12}$: from 2.174 to 2.262 {\AA}, $\sim4\%$ expansion). In this regime, adhesion can be large even if $\rho_{red}$ computed within a fixed interfacial region is comparatively low, because part of the charge rearrangement is spatially distributed into the carbon slab rather than being concentrated only at the immediate contact. Overall, these two outliers emphasize that $\rho_{red}$ quantifies the magnitude of charge rearrangement within a chosen interfacial region, but the adhesion also depends on whether that rearrangement reflects long-range metallic screening (as in Al) or efficient bond-forming chemisorption with structural accommodation (as in Ti).

\section{CONCLUSIONS}
In this work, a high-throughput \textit{ab initio} investigation of diamond/metal interfaces has been carried out, systematically analyzing the adhesion of 60 combinations involving a broad range of metallic substrates and four low-index diamond terminations (C(111), C(111)-rec, C(110), C(100)), representative of the different types of bonds in DLC coatings. The study employs an automated workflow implemented within the TribChem software, which ensures a consistent and reproducible protocol for the construction and characterization of solid interfaces. In particular, the interface supercells were built using the algorithm developed by Zur, which identifies the smallest common supercell between two crystalline surfaces while minimizing both lattice mismatch and angular distortion. All interfaces were constructed with a metallic strain below 3$\%$, ensuring that the calculated adhesion energies are not affected by unphysical distortions.\par
The results show that adhesion is primarily governed by the intrinsic reactivity of the metallic component and by the type of carbon hybridization present at the surface of the diamond. Carbide-forming metals such as Cr, Ti, W and V exhibit the strongest adhesion, in agreement with their experimentally recognized role as interlayer materials that enhance DLC adhesion to metallic substrates. In contrast, noble and weakly reactive metals (Ag, Au, Cu, Zn, Mg) display poor adhesion due to their filled valence shells and low surface energies. Among the diamond terminations, the C(111) surface with sp$^3$ bonding shows the highest adhesion, whereas surface graphitization, as observed in C(111)-rec, considerably weakens interfacial bonding.\par
A strong correlation between adhesion and the geometric mean of the constituent surface energies at fixed diamond termination, $E_{adh}\simeq\sqrt{\gamma_1\gamma_2}$, confirms that adhesion strength can be rationalized from intrinsic surface properties. This provides an efficient descriptor for fast materials screening, without the need to simulate computationally expensive interface models.
\\By comparing the work of separation ($W_{sep}=-E_{adh}$) of the diamond/metal interface with the cohesive energy of the corresponding bulk metal, it is possible to predict the likely fracture location under tensile load. Interfaces involving C(111)-rec consistently exhibit fracture at the interface, reflecting the weak bonding due to sp$^2$ terminations. For metals with low cohesive energies, such as Mg, Zn and Al, fracture can occur within the metal bulk when in contact with highly reactive diamond surfaces, indicating that the interface is stronger than the metallic cohesion. Conversely, transition metals with partially filled d-shells (e.g., W, Cr, Mo, Fe) exhibit high bulk cohesion, preventing failure from taking place within the metal phase.\par
Analysis of the radial distribution function provides insights on the structural response of diamond surface reconstructions upon contact with metals. In C(111)-rec, the peak at approximately 1.44 {\AA}, corresponding to the zig-zag surface chains, is largely preserved in the presence of weakly interacting metals such as Au and Ag, indicating minimal perturbation of the sp$^2$ network. In contrast, Cr, Al, Cu and Pt induce significant dimerization and partial rehybridization of the surface sp$^2$ bonds. In general, reactive metals like W, Cr, Mo, Fe, V and Ti are able to reduce the number of surface sp$^2$ bonds by up to 50$\%$. Hence, it can be concluded that strong perturbation and reduction in the number of surface sp$^2$ bonds in diamond coatings directly enhance adhesion, explaining why these metals are commonly used as interlayers to improve bonding between DLC films and metallic substrates.\\
The structural modifications are mirrored in an electronic rearrangement at the interface. Volumetric charge density differences show that metals inducing stronger structural perturbations promote more pronounced charge rearrangement, while weaker interacting metals cause only minor redistribution. Quantitatively, the charge redistribution descriptor, $\rho_{\rm red}$, remains strongly correlated with adhesion, confirming that electronic charge redistribution upon interface formation is a good predictor of adhesion strength, even in the presence of semiconductor-metal bonding. \par
In summary, by focusing on ideal crystalline interfaces, this work aimed at establishing physically grounded and broadly applicable correlations linking adhesion to intrinsic structural and electronic descriptors, without resorting to system specific models and by exploiting the accuracy of first principles calculations. In the future, machine learning approaches could leverage these insights to address more realistic defective or amorphous interfaces and larger length/time scales.

\begin{acknowledgments}
These results are part of the "Advancing Solid Interface and Lubricants by First Principles Material Design (SLIDE)" project that has received funding from the European Research Council (ERC) under the European Union's Horizon 2020 research and innovation program (Grant agreement No. 865633). The authors acknowledge funding by the European Union - NextGenerationEU (National Sustainable Mobility Center CN00000023, Italian Ministry of University and Research Decree n. 1033 - 17/06/2022, Spoke 11 - Innovative Materials $\&$ Lightweighting) and by ICSC-Centro Nazionale di Ricerca in High Performance Computing,Big Data and Quantum Computing, funded by EuropeanUnion-NextGenerationEU. The opinions expressed are those of the authors only and should not be considered as representative of the European Union or the European Commission’s official position. Neither the European Union nor the European Commission can be held responsible for them. We acknowledge the CINECA award under the ISCRA initiative, for the availability of high-performance computing resources and support.
\end{acknowledgments}

%\nocite{*}

\bibliography{bib}% Produces the bibliography via BibTeX.

\end{document}